\documentstyle[twoside,epsf]{article}

\input ibvs2.sty

\begin{document}

\IBVShead{6175}{27 June 2016}

\IBVStitletl{Metal-rich or misclassified?}{The case of four RR Lyrae stars}

\IBVSauth{Moln\'ar, L.$^{1}$; Juh\'asz$^{1,2}$, \'A. L.; Plachy, E.$^{1}$; Szab\'o, R.$^1$}

\IBVSinst{Konkoly Observatory, Research Centre for Astronomy and Earth Sciences, Konkoly Thege Mikl\'os \'ut 15-17, H-1121 Budapest, Hungary}
\IBVSinst{Department of Astronomy, E\"otv\"os Lor\'and University, P\'azm\'any P\'eter s\'et\'any 1/a, H-1117 Budapest, Hungary}
\IBVSinsto{e-mail: molnar.laszlo@csfk.mta.hu}

\IBVSabs{We analysed the light curve of four, apparently extremely metal-rich fundamenta-mode }
\IBVSabs{RR Lyrae stars. We identified two stars, MT Tel and ASAS J091803-3022.6 as RRc }
\IBVSabs{(first-overtone) pulsators that were misclassified as RRab ones in the ASAS survey. In the }
\IBVSabs{case of the other two stars, V397 Gem and ASAS J075127-4136.3, we could not decide }
\IBVSabs{conclusively, as they are outliers in the period-Fourier-coefficient space from the loci }
\IBVSabs{of both classes, but their photometric metallicities also favour the RRc classification. }

\begintext
\section{Introduction}
RR Lyrae stars are old, usually metal-poor, population II stars, currently evolving on the horizontal branch, and crossing the classical instability strip. Most of them, the RRab stars, pulsate in the fundamental mode and have well recognisable light curves, with a very short and steep ascending branch and a long descending branch that often, but not always, exhibits a strong bump feature before reaching minimum light. First-overtone pulsators, members the RRc subclass, display a comparatively more sinusoidal light curve that is usually less asymmetric, and often features a notable depression, or hump, before maximum light. (Quasi-continuous observations of space photometric missions provided numerous textbook examples of RR Lyrae light curves: we refer the reader to the works by Benk\H{o} et~al.\ 2010, 2014, Moskalik et~al.\ 2015, Nemec et~al.\ 2011, Szab\'o et~al.\ 2014, and references therein.)

The light curve shape of RR Lyrae stars strongly depends on the physical parameters of the star, therefore photometric data can be exploited to derive those properties. Jurcsik \& Kovacs (1996, JK96 hereafter) derived a formula to calculate the [Fe/H] indices of RRab stars from the pulsation period and the $\phi_{31}$ Fourier coefficients, defined as the $\phi_{31} = \phi_3-3\,\phi_1$ relative phase difference by Simon \& Lee (1981). A similar relation was determined for RRc stars by Morgan et~al.\ (2007), and later modified by Nemec et~al.\ (2013), who also provided an updated formula for RRab stars observed in the passband of the \textit{Kepler} space telescope. We note that the accuracy of the JK96 formula is limited for very low- and high-metallicity stars (see, e.g., Nemec et~al.\ 2013).

While most RR Lyrae stars are metal-poor, with negative [Fe/H] indices (see, e.g.\ Feast et~al.\ 2008), some may reach metallicities close to solar, such as AV Peg (--0.08), RW TrA (--0.13) from the HIPPARCOS sample (Feast et~al.\ 2008) or V839 Cyg ($-0.05\pm0.14$) and V784 Cyg ($-0.05\pm0.10$) in the original \textit{Kepler} sample (Nemec et~al.\ 2013). Another interesting example is the short-period, high-metallicity star TV Lib (with  ($P$ = 0.26962~d, and [Fe/H] = --0.17) that nevertheless displays very characteristic RRab light variations (Clube et~al.\ 1969, Kov\'acs 2005). 

During the target search for the K2 mission of the \textit{Kepler} space telescope (Howell et~al.\ 2014; Plachy et~al.\ 2016) we encountered one star, V397 Gem, whose photometric metallicity appeared to be extremely high, [Fe/H] = 0.42, based on the NSVS light curve (Northern Sky Variability Survey, Hoffmann et~al.\ 2009) and the original formula of JK96. V397 Gem also displayed a very short pulsation period of $P = 0.294$ d, considering that it was classified as an RRab star by Wils et~al.\ (2006). After a literature search, we found three more examples in the measurements of the All-Sky Automated Sky Survey (ASAS), based on the work of Szczygie\l{} et~al.\ (2009). All three stars were classified as RRab ones, have short periods (between 0.3--0.35 d) and [Fe/H] indices consistently above +0.5, based on methods of both JK96 and Sandage (2004). The folded light curves of the four stars are displayed in Fig.\ 1. 

The apparently extremely high metallicity suggests that these stars cannot be RRab stars, and the calculated values are simply erroneous. Alternate possibilities include high-amplitude $\delta$ Scuti stars; their rare, high-mass variants, the AC And-type pulsators; and RRc stars with unusually asymmetric, sawtooth-like light curves that mislead the (semi-)automated classification schemes of the surveys. In this paper we examine whether these objects could be misclassified RRc stars.

\IBVSfig{11cm}{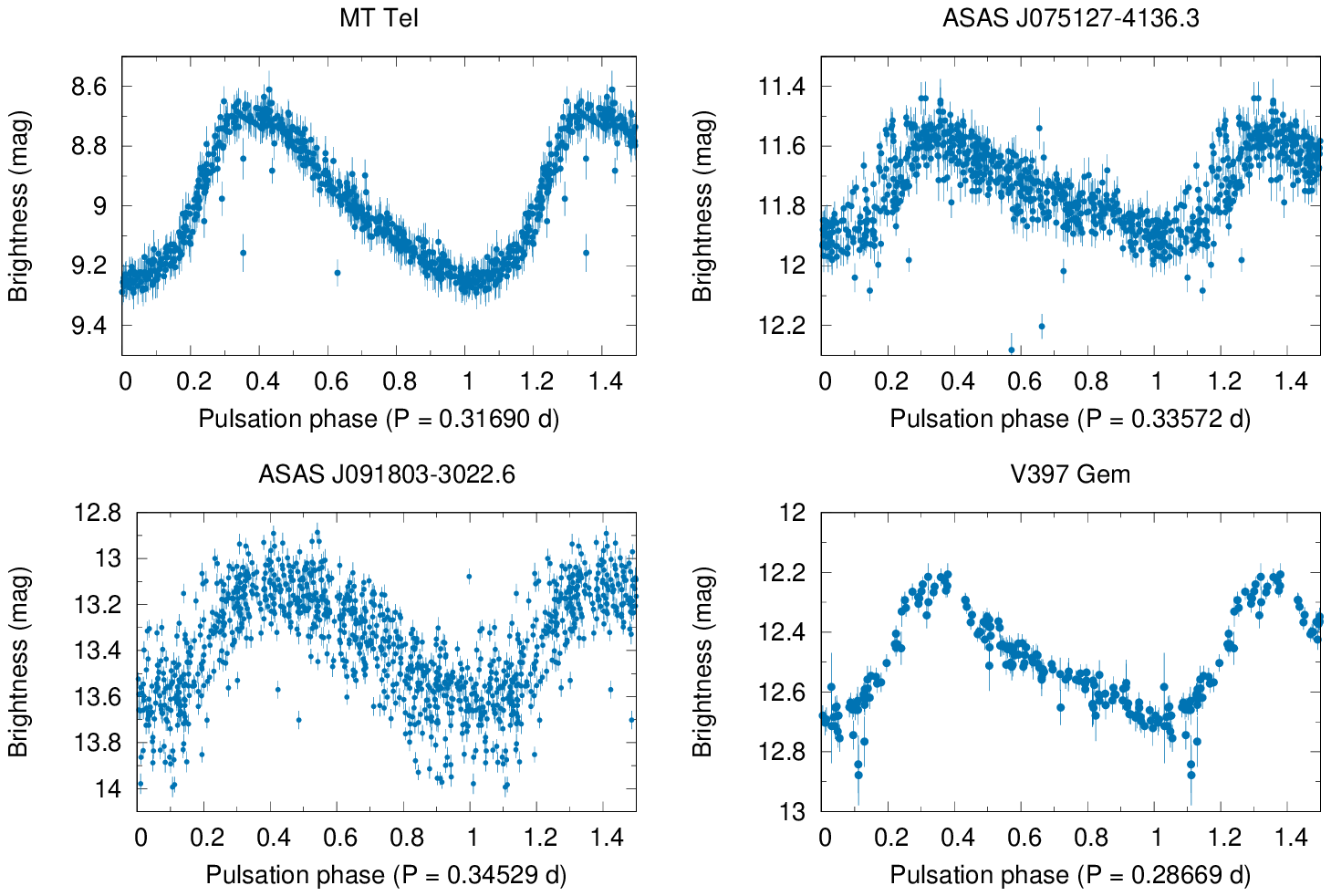}{Folded light curves of the four RR Lyrae stars. Data were obtained by the NSVS survey for V397 Gem, and by the ASAS survey for the other three stars.}
\IBVSfigKey{6175-f1.eps}{MT Tel; V397 Gem; ASAS J075127-4136.3; ASAS J091803-3022.6}{light curve}

\section{Photometric metallicities}
As a first step, we recalculated the photometric [Fe/H] incides of the stars, using equation (4) of Nemec et~al.\ (2013) which is an adjusted version of the Morgan et~al.\ (2007) relation, with the addition of four RRc stars observed by \textit{Kepler} during the original mission. In all four cases we derived indices typical of RR Lyrae stars, between --1.70 and --2.14. One of the four stars, MT Tel was actually known to be a first-overtone star, but even Przybylski (1967) mentioned in the discovery paper that the star displayed an unusually asymmetric light curve for an RRc star. Our value, [Fe/H] = --1.97, agrees relatively well with the one derived by Feast et~al.\ (2008) who calculated it to be --1.85. These findings suggest (although do not prove) that the stars are potentially misclassified in the works of Wils et~al.\ (2006) and Szczygie\l{} et~al.\ (2009), rather than being extremely metal-rich ones.

\vskip 5mm
\begin{center}
\begin{table}[h]
\caption{Properties of the four RR Lyrae stars. [Fe/H]$_{\rm RRab}$ and [Fe/H]$_{\rm RRc}$ columns are the photometric metallicity values obtained from the equations for fundamental-mode and first-overtone stars, respectively.}
\vskip 5mm
\begin{tabular}{lcccccc}
\hline
Name & RA$_{2000}$ & Dec$_{2000}$ & $V$ & Period & [Fe/H]$_{\rm RRab}$ & [Fe/H]$_{\rm RRc}$  \\
~ & h:m:s & d:m:s & mag & d &~ &~\\
\hline
V397 Gem & 06:22:44.3 & +18:31:53.4 & 12.1 & 0.28669 & 0.42  & $-$1.70\\ 
MT Tel & 19:02:12.0 & -46:39:11.9 & 8.94 & 0.31690 & 0.66/1.04 & $-$1.97\\
J075127--4136.3 & 07:51:27.4 & --41:36:17.9 & 11.86 & 0.33572 & 0.95/1.13 & $-$2.11 \\
J091803--3022.6 & 	09:17:59.6 & --30:23:34.0 & 12.06 & 0.34529 & 1.51/0.85 & $-$2.14\\
\hline
\end{tabular}
\end{table}
\end{center}

\section{Fourier coefficients}
We calculated the various Fourier coefficients of the four stars and plotted them against the OGLE-IV bulge sample and a collection of RRab stars observed by the ASAS and SuperWASP surveys in Fig.\ 2 (Skarka 2014). The OGLE data, published by Soszy\'nski et~al.\ (2014), were collected in the \textit{I} band, and we converted their coefficients to \textit{V} with the conversion formulae of Morgan et~al.\ (1998). We note that the \textit{c} indices indicate that here the $\phi_{21}$ and $\phi_{31}$ phase differences were converted to cosine-based Fourier parameters: $\phi_{21}^c = \phi_{21}^s + \pi/2$ and $\phi_{31}^c=\phi_{31}^s-\pi$. 

We also highlighted three known high-metallicity stars in the plot with black circles for comparison with the stars we found. The two longer-period ones, RW TrA and AV Peg, fall into the RRab loci in all four plots. The third one, TV Lib, has already been known as a very peculiar star, especially because of its very short period, and it is an outlier in three out of four panels here too.  

Based on the positions of the four stars, we can summarise tour findings as follows:

\begin{itemize}
\item V397 Gem, the star with the shortest period out of the four, falls into the region that is populated, sparsely, by both classes. Based on the Fourier coefficients alone, the classification of this star is still uncertain. Together with the photometric metallicity values, the RRc classification is more plausible, but not definitive. 
\item  MT Tel is also close to the overlapping region in three out of four panels, but the low $\phi_{31}^c$ confirms that it is an RRc star, albeit an unusual one, as it has been already established in previous works.
\item  ASAS J075127--4136.3 is another star that falls into the overlapping either-or region in the first three panels. The photometric metallicity and the $\phi_{31}^c$ value together suggest that the star is likely a first overtone pulsator rather than a fundamental-mode one, but our classification is uncertain. 
\item ASAS J091803--3022.6, the star with the longest period and lowest $R_{21}$ ratio, falls squarely into the RRc loci, confirming that it is indeed a first-overtone star. 
\end{itemize}

We note that the classification of the OGLE-IV observations was based on fitting light curve templates typical of the RRab and RRc classes to the data. Hence it is entirely possible that some stars with highly asymmetric, sawtooth-like, short-period variations, similar to the ones shown in Fig.~1, were identified as RRab stars in that sample too. 

\IBVSfig{11.5cm}{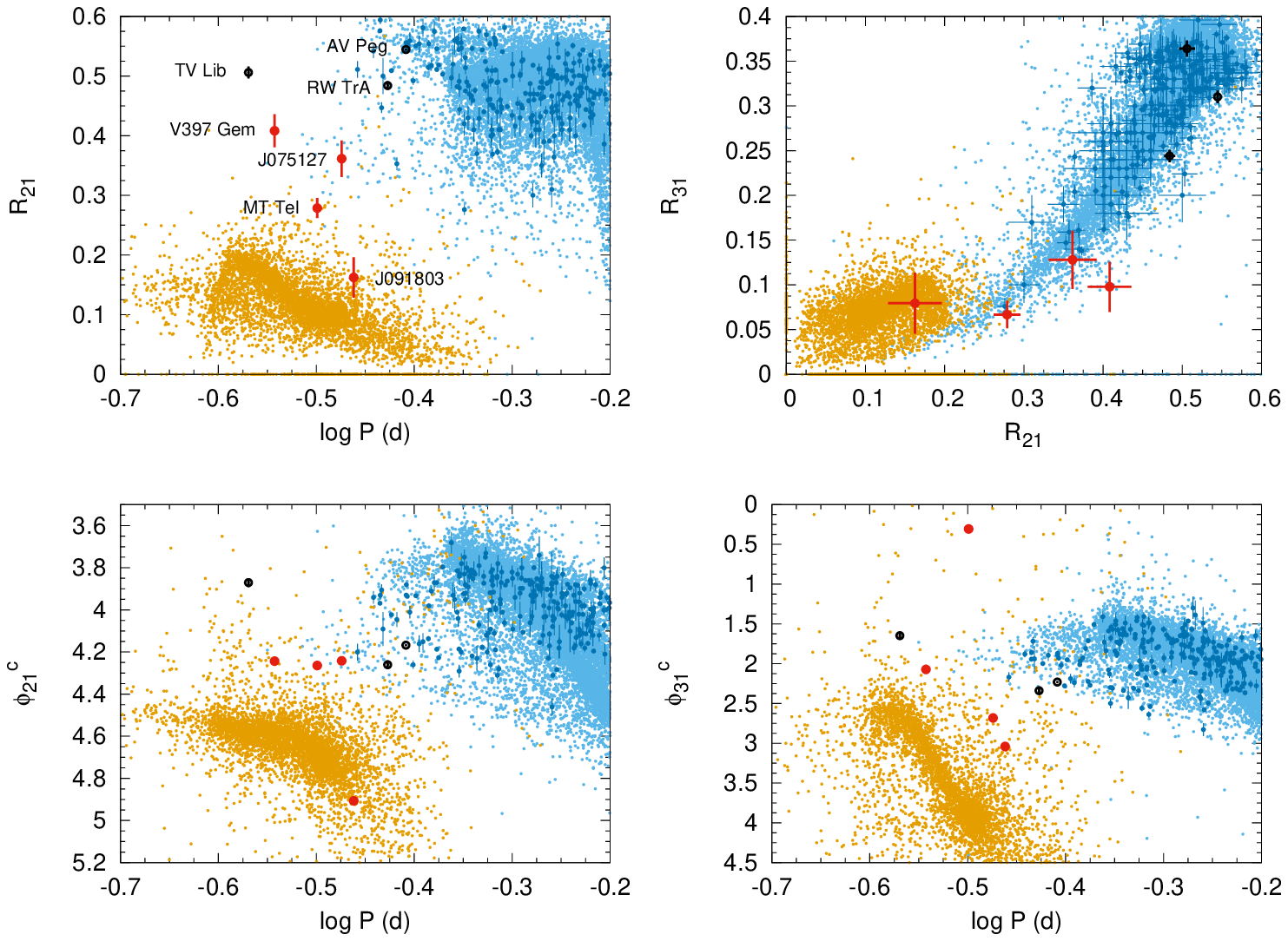}{Fourier coefficients of bulge RRab (light blue) and RRc (orange) stars from OGLE-IV, field RRab stars (dark blue), and the four stars from this paper (red). Known high-metallicity RRab stars are marked with black circles. Uncertainties for the OGLE-IV data were not provided. }
\IBVSfigKey{6175-f2.eps}{MT Tel; V397 Gem; ASAS J075127-4136.3; ASAS J091803-3022.6;TV Lib; RW TrA; AV Peg}{other}

\section{Conclusions}
The analysis of these four stars highlights the limits of the photometric methods in variable star classification. Out of the four stars we examined, MT Tel was already known to be an RRc star, yet the classification scheme of ASAS labeled it as a fundamental-mode star. We showed that another star, ASAS J091803-3022.6 was also misclassified and is in fact an RRc star, but in this case the error be plausibly attributed to the low photometric quality of the ASAS data. In both cases the reclassification also provided us with [Fe/H] indices that are common for RR Lyrae stars, instead of the apparently extremely high (and therefore likely erroneous) values calculated by Szczygie\l{} et~al.\ (2009). 

The case of the other two stars, V397 Gem and ASAS J075127-4136.3, is less simple. The period--Fourier-coefficient parameter space contains multiple outlier points, from both subclasses: V397 Gem and ASAS J075127-4136.3 are prime examples of these stars (as is TV Lib). While we prefer the RRc classification for both stars, spectroscopic observations will be needed to decide between the two, very different [Fe/H] indices we obtained from photometry. We also note that in the case of V397 Gem, high-precision photometry from the \textit{Kepler} space telescope might be able to help, as the low-amplitude additional mode content of RRc and RRab stars are different (see, e.g., Benk\H{o} et~al.\ 2014, Moskalik et~al.\ 2015, Moln\'ar 2016) but the investigation of the K2 data will be part of a separate study. 

\vskip 4 mm

\emph{Acknowledgements:} L.~M. was supported by the J\'anos Bolyai Research Scholarship and the INKP 2015/1 travel grant  of the Hungarian Academy of Sciences. This work has been supported by the NKFIH K-115709 and PD-116175 grants of the Hungarian National Research, Development and Innovation Office. The research leading to these results has received funding from the ESA PECS Contract No.\ 4000110889/14/NL/NDe and the European Community's Seventh Framework Programme  (FP7/2007-2013) under grant agreements no. 269194 (IRSES/ASK) and no. 312844 (SPACEINN). This research has made use of the SIMBAD database, operated at CDS, Strasbourg (France), and NASA's Astrophysics Data System Bibliographic Services.

\vskip -2 mm

\references

Benk\H{o}, J.\ M., et al., 2010, {\it MNRAS}, {\bf 409}, 1585

Benk\H{o}, J.\ M., Plachy, E., Szab\'o, R., Moln\'ar, L., Koll\'ath, Z., 2014, {\it ApJS}, \textbf{213}, 31

Clube, S.~V.~M., Evans, D.~S., Jones, D.~H.~P., 1969, \textit{Mem.\ RAS}, \textbf{72}, 101

Feast, M.~W., Laney, C.~D., Kinman, T.~D., van Leeuwen, F., Whitelock, P.~A., 2008, \textit{MNRAS}, \textbf{386}, 2115

Hoffman, D.~I., Harrison, T.~E., McNamara, B.~J., 2009, \textit{AJ}, \textbf{138}, 466

Howell, S.\ B., et al., 2014, {\it PASP}, \textbf{126}, 398

Jurcsik, J., Kovacs, G., 1996, \textit{A\&A}, \textbf{312}, 111

Kov\'acs, G., 2005, \textit{A\&A}, \textbf{438}, 227

Moln\'ar, L., 2016, \textit{CoKon}, \textbf{105}, 11

Morgan, S.~M., Simet, M., Bargenquast, S., 1998, \textit{AcA}, \textbf{48}, 341

Morgan, S.~M., Wahl, J.~N., Wieckhorst, R.~M., 2007, \textit{MNRAS}, \textbf{374}, 1421

Moskalik, P. A., 2015, \textit{MNRAS}, \textbf{447}, 2348

Nemec, J.~M., et al., 2011, \textit{MNRAS}, \textbf{417}, 1022

Nemec, J.~M., et al., 2013, \textit{ApJ}, \textbf{773}, 181

Plachy, E., Moln\'ar, L., Szab\'o, R., Kolenberg, K., B\'anyai, E., 2016, \textit{CoKon}, \textbf{105}, 19

Przybylski, A., 1967, \textit{MNRAS}, \textbf{136}, 185

Sandage, A., 2004, \textit{AJ}, \textbf{128}, 858

Simon, N.~R., Lee, A.~S., 1981, \textit{ApJ}, \textbf{248}, 291

Skarka, M., 2014, \textit{MNRAS}, \textbf{445}, 1584

Soszy\'nski, I., et al., 2014, \textit{AcA}, \textbf{64}, 177

Szab\'o, R., et al., 2014, {\it A\&A}, \textbf{570}, 100

Szczygie\l{}, D.~M., Pojman\'nski, G., Pilecki, B., 2009, \textit{AcA}, \textbf{59}, 137

Wils, P., Lloyd, C., Bernhard, K., 2006, \textit{MNRAS}, \textbf{368}, 1757

\endreferences

\end{document}